\documentclass[aps,showpacs,showkeys,twocolumn]{revtex4}

\usepackage{amssymb}
\usepackage{bm}
\begin{document}

\def\a{\alpha}
\def\b{\beta}
\def\l{\lambda}
\def\e{\epsilon}
\def\p{\partial}
\def\m{\mu}
\def\n{\nu}
\def\t{\tau}
\def\th{\theta}
\def\s{\sigma}
\def\g{\gamma}
\def\o{\omega}
\def\r{\rho}
\def\half{\frac{1}{2}}
\def\hatt{{\hat t}}
\def\hatx{{\hat x}}
\def\hatp{{\hat p}}
\def\hatX{{\hat X}}
\def\hatY{{\hat Y}}
\def\hatP{{\hat P}}
\def\hatth{{\hat \theta}}
\def\hatta{{\hat \tau}}
\def\hatrh{{\hat \rho}}
\def\hatva{{\hat \varphi}}
\def\barx{{\bar x}}
\def\bary{{\bar y}}
\def\barz{{\bar z}}
\def\baro{{\bar \omega}}
\def\sp{\sigma^\prime}
\def\nn{\nonumber}
\def\cb{{\cal B}}
\def\2pap{2\pi\alpha^\prime}
\def\wideA{\widehat{A}}
\def\wideF{\widehat{F}}
\def\beq{\begin{eqnarray}}
 \def\eeq{\end{eqnarray}}
 \def\4pap{4\pi\a^\prime}
 \def\xp{{x^\prime}}
 \def\sp{{\s^\prime}}
 \def\ap{{\a^\prime}}
 \def\tp{{\t^\prime}}
 \def\zp{{z^\prime}}
 \def\xpp{x^{\prime\prime}}
 \def\xppp{x^{\prime\prime\prime}}
 \def\barxp{{\bar x}^\prime}
 \def\barxpp{{\bar x}^{\prime\prime}}
 \def\barxppp{{\bar x}^{\prime\prime\prime}}
 \def\barchi{{\bar \chi}}
 \def\baro{{\bar \omega}}
 \def\bpsi{{\bar \psi}}
 \def\barg{{\bar g}}
 \def\barz{{\bar z}}
 \def\bareta{{\bar \eta}}
 \def\ta{{\tilde \a}}
 \def\tb{{\tilde \b}}
 \def\tc{{\tilde c}}
 \def\tz{{\tilde z}}
 \def\tJ{{\tilde J}}
 \def\tpsi{\tilde{\psi}}
 \def\tal{{\tilde \alpha}}
 \def\tbe{{\tilde \beta}}
 \def\tga{{\tilde \gamma}}
 \def\tchi{{\tilde{\chi}}}
 \def\barth{{\bar \theta}}
 \def\bareta{{\bar \eta}}
 \def\barom{{\bar \omega}}
\setcounter{page}{1}
\title[]{Dissipative Hofstadter Model at the Magic Points and\\ 
Critical Boundary Sine-Gordon Model}

\author{Seungmuk Ji}
\affiliation{
Department of Physics, Kangwon National University, Chuncheon 200-701
Korea and\\
Korea Research Institute of Standards and Science,
P.O.Box 102, Yuseong, Daejeon 305-600, Korea}
\author{Ja-Yong Koo}
\affiliation{
Korea Research Institute of Standards and Science,
P.O.Box 102, Yuseong, Daejeon 305-600, Korea}
\author{Taejin Lee$^*$}
\affiliation{
Department of Physics, Kangwon National University, Chuncheon 200-701
Korea and\\ Pacific
Institute for Theoretical Physics, University of British Columbia, 
6224 Agricultural Road, Vancouver, British Columbia V6T 1Z1, Canada}

\email{taejin@kangwon.ac.kr}

\begin{abstract}
The dissipative Hofstadter model describes quantum particles moving 
in two dimensions subject to a uniform magnetic field, a periodic 
potential and a dissipative force. We discuss the dissipative 
Hofstadter model in the framework of the boundary state 
formulation in string theory and construct exact boundary states 
for the model at the magic points by using the fermion representation. The dissipative Hofstadter 
model at magic points is shown to be equivalent to the critical boundary sine-Gordon model. 
\end{abstract}



\maketitle

\begin{center}
{\bf I. INTRODUCTION}
\end{center}

The quantum mechanical description of dissipation has been one of 
the outstanding problems in theoretical physics. Since the well-known dissipative or
frictional force term in the classical equation of motion cannot be driven 
from a local action, it has been a conundrum how to quantize the dissipative 
system for a long time. Caldeira and Leggett \cite{caldeira83ann,caldeira83phy}
provided a proper answer to this problem
by coupling a bath or environment which consists of an infinite 
number of harmonic oscillators to the system. Assuming that the interaction between the 
bath and the system is linear and imposing the Ohmic condition for the spectral function 
of the oscillator frequencies, they found that the effective friction term could be
generated in the equation of motion. 
In the quantum theory the interaction with the bath produces a non-local effective
interaction. This model for the dissipative system is now called the
Caldeira-Leggett model. 

Since the effective interaction is nonlinear, the dissipative quantum system 
exhibits various kinds of phase transition, unlike the quantum 
mechanical systems with local interactions only. One of the 
interesting dissipative systems, which has been extensively studied 
for its novel phase diagram \cite{callan91,callan93,freed93}, is 
the dissipative Hofstadter model.
In condensed matter physics, the dissipative Hofstadter model has a wide range of 
applications which include Josephson junction arrays \cite{larkin,fazio,sodano},
the Kondo problem \cite{Affleck:1990iv,Affleck:1990by,hong}, 
the study of one-dimensional
conductors \cite{Kane:1992}, tunnelling between Hall edge states
\cite{kane2}, and junctions of quantum wires~\cite{Oshikawa:2005fh}.
The dissipative Hofstadter model also has an important application
in string theory, since the boundary (conformal) field theories describing the
dissipative Hofstadter model are solutions of classical open string field 
theory~\cite{Witten:1992qy,Shatashvili:1993kk,Shatashvili:1993ps}.
In particular, a marginal, periodic boundary interaction which is
termed the ``rolling tachyon'' \cite{Sen:2002nu,senreview,kim,lyi} 
gives a description of the process of
tachyon condensation \cite{lee0105,lee01,lee01052,lee0109}  
in string theories with unstable D-branes. The conformal field theory 
for the rolling tachyon
corresponds to the Schmid model \cite{schmid,fisher,guinea}
at the critical point (also called the magic 
point), which is the dissipative Hofstadter model in the absence 
of the magnetic field.

The dissipative Hofstadter model is a one dimensional quantum 
mechanical system which has time dimension only.  
In the string theory, the time of the dissipative Hofstadter 
model is mapped onto the boundary of the disk diagram and the 
bulk degrees of freedom of the string play the role of the bath to 
produce the dissipative non-local interaction at the boundary. 
As is wellknown, the string theory on a disk can be discussed best in 
the framework of the boundary state formulation. Thus, the 
boundary state formulation would be the most suitable framework to 
discuss the dissipative Hofstadter model, at least at the magic 
points where the boundary interaction becomes exactly marginal.
Along this line the Schmid model at the critical point was first discussed 
by Callan and Thorlacius \cite{callan90} and later the dissipative Hofstadter 
model at the magic points by Maldacena {\it et al}.   

One of the authors has recently discussed the rolling tachyon
boundary state by fermionizing the rolling tachyon boundary
conformal field theory and then constructed the exact boundary states by using
fermion variables~\cite{Hassel}. Extending the previous work, 
we shall discuss the dissipative Hofstadter model at the magic 
points in the framework of the boundary state formulation in the 
string theory using the fermion representation. The dissipative 
Hofstadter model at the magic points will be shown to be 
equivalent to the boundary sine-Gordon model and the mobility 
will be calculated exactly by using the fermion representation of 
the boundary state for the sine-Gordon model.

\begin{center}
{\bf II. The Schmid Model}
\end{center}

In the absence of the magnetic field, the dissipative Hofstadter 
model reduces to the Schmid model, of which the action is given as 
follows:
\beq 
S_{SM} &=& \frac{\eta}{4\pi \hbar} \int^{T/2}_{-T/2} dt dt^\prime 
\frac{\left(X(t) - X(t^\prime)\right)^2}{(t-t^\prime)^2} 
+ \nn\\
&&  \frac{M}{2\hbar} \int^{T/2}_{-T/2} dt \dot X^2
- \frac{V_0}{\hbar} \int^{T/2}_{-T/2} dt \cos \frac{2\pi X}{a}. 
\eeq
The first term is responsible for the dissipation, and the second 
term is the usual kinetic term for a particle with mass $M$. The 
third term denotes the periodic potential. The period of 
time is to be $T$. 
If we wish to describe the system in real time, we
may take the limit where $T \rightarrow \infty$. Since we are only 
interested in the long-time behavior of the system, we may ignore
the kinetic term, which only plays a role of regulator in the 
long-time analysis. 

Mapping the Schmid model to the string theory on a disk
begins with identifying the time as the boundary parameter $\sigma$ 
in string theory and scaling the field variable $X$:
\beq
t = \frac{T}{2\pi} \s, \quad X \rightarrow \frac{a}{2\pi} X.
\eeq
Then, the action for the Schmid model reads as 
\beq
S_{SM} &=& \frac{\eta}{4\pi \hbar} \left(\frac{a}{2\pi}\right)^2 
\int^{\pi}_{-\pi} d\s d\s^\prime 
\frac{\left(X(\s) - X(\s^\prime)\right)^2}{(\s-\s^\prime)^2} \nn\\
&& - \frac{V_0}{\hbar} \frac{T}{2\pi} \int^{\pi}_{-\pi} d\s \half
\left(e^{iX} + e^{-iX} \right). 
\eeq
This action precisely coincides with the boundary effective action 
for the bosonic string subject to a boundary periodic potential
on a disk with a boundary condition $X(\t=0,\s) = X(\s)$,
\beq
e^{-S_{SM}} &=& \int D[X] \exp\Biggl[ - \frac{1}{4\pi \ap}
\int d\t d\s \p_\a X \p^\a X \nn\\
&&  + \frac{g}{2}
\int d\s \left(e^{iX}+ e^{-iX} \right)\Biggr]. 
\eeq 
Here, we identify the physical parameters of the two theories as 
\beq
\frac{\eta}{4\pi \hbar} \left(\frac{a}{2\pi}\right)^2 
= \frac{1}{8\pi^2 \ap}, \quad 
-\frac{V_0}{\hbar} \frac{T}{2\pi} = g.
\eeq
The action for the Schmid model also 
appears in the boundary state for the bosonic string 
\beq
|B \rangle &=& \int D[X] \exp\left(g \int d\s \cos X \right) |X \rangle, \\
|X \rangle &=& e^{- \frac{1}{8 \pi^2 \ap}
\int^{\pi}_{-\pi} d\s d\s^\prime 
\frac{\left(X(\s) - X(\s^\prime)\right)^2}{(\s-\s^\prime)^2} + 
\dots }|0 \rangle. \nn  
\eeq
Thus, 
\beq \label{effective}
\langle 0 |B\rangle = \int D[X] \exp \left(-S_{SM}[X] \right).
\eeq

The critical point for the Schmid model corresponds to
the point where $\ap =1$, {\it i.e.}, 
the boundary periodic potential becomes marginal. As you may 
notice, the Schmid model at the critical point is nothing but the 
conformal field theory for the rolling tachyon of the 
full-S-brane.
As discussed in ref. \cite{Hassel}, at the critical point the 
conformal theory can be better understood in its fermion 
representation \cite{pol,Lee:2005ge}; the fermionic representation has the advantage that the 
boundary interaction becomes a simple fermion current operator which 
is bilinear in fermion fields. As a result, one can easily construct 
an exact boundary state for the theory. In order to fermionize the 
system, we need to introduce an auxiliary free boson $Y$ which obeys a Dirichlet boundary condition on all boundaries. This allows us to form 
the bosonic variables 
\begin{equation}\phi_1 = \frac{1}{\sqrt{2}}\left( X+Y\right),~~
\phi_2=\frac{1}{\sqrt{2}}\left(X-Y\right),\label{variables}\end{equation}
with which the mapping to fermions is defined:
\begin{eqnarray}\label{fermionization1}
\psi_{1L}(z)&=&\zeta_{1L}:e^{-\sqrt{2}i\phi_{1L}(z)}:,\nn\\
\psi_{2L}(z)&=&\zeta_{2L}:e^{\sqrt{2}i\phi_{2L}(z)}:, \nn\\
\psi_{1R}(\bar z)&=&\zeta_{1R}:e^{\sqrt{2}i\phi_{1R}(\bar 
z)}:,\nn\\
\psi_{2R}(\bar z)&=& \zeta_{2R} :e^{-\sqrt{2}i\phi_{2R}(\bar 
z)}:,
\end{eqnarray}
where $\zeta_{aL/R}$ are co-cycles, ensuring the anti-commutation
relations between the fermion operators.
In terms of the fermion fields, the string action for the Schmid 
model is given by
\beq
&&\int d\t d\s L \nn\\
&& = \int \frac{d\tau d\sigma}{2\pi} ~
\left[ \psi_L^{\dagger}\left(\partial_\tau+ \partial_\sigma\right)
\psi_L + \psi_R^{\dagger}\left(\partial_\tau-
i\partial_\sigma\right)\psi_R\right] \nn\\
&& + ig\pi \int \frac{d\s}{2\pi}
\Bigl[\psi^\dagger_L \left(\frac{1+\s^3}{2}\right) \psi_R + \nn\\
&&~~~~~~~~~~~~~~~~~~~~~~~~~~~~~~~
\psi^\dagger_L \left(\frac{1- \s^3}{2}\right) \psi_R \Bigr],
\eeq
where $\psi_L = (\psi_{1L}, \psi_{2L})^t$,  
$\psi_R = (\psi_{1R}, \psi_{2R})^t$.
We quote from ref. \cite{Hassel} the explicit expression of the 
boundary state. This is in the $NS$-sector
\beq \label{nsboundarystate}
|BD\rangle_{NS} 
=\prod_{r=\frac{1}{2}}^\infty
e^{\psi_{-r}^\dagger U^{-1}i\sigma^1\tilde\psi_{-r} -
\tilde\psi^{\dagger}_{-r}i\sigma^1U\psi_{-r}}|0\rangle,
\eeq
and in the R-sector 
\beq \label{rboundarystate}
|BD\rangle_R 
&=&\prod_{n=1}^\infty  \exp\left[
\psi_{-n}^\dagger U^{-1}i\sigma^1\tilde\psi_{-n} -
\tilde\psi^{\dagger}_{-n}i\sigma^1U\psi_{-n}\right]
\nn\\
&&~~~\exp\left[\psi_0^\dagger
U^{-1}i\sigma^1\tilde\psi_0\right]~|-+-+\rangle, 
\eeq
where
\beq
U = \left[\begin{array}{cc}
  \sqrt{1-\pi^2 g^2} & -i\pi g \\
  -i\pi g  & \sqrt{1-\pi^2 g^2}  
\end{array} \right].
\eeq
Note that the matrix $U$ is not unitary if the coupling exceeds a 
critical strength, $|g|>1/\pi$. (For the notation, the reader 
should refer to ref. \cite{Hassel}.) 

\begin{center}
{\bf III. The Dissipative Hofstadter model at Magic Points}
\end{center}

If we turn on the magnetic field, the Schmid model becomes the 
dissipative Hofstadter model, of which the action is given by
\beq 
S_{DHM} &=& \frac{\eta}{4\pi \hbar} \int^{T/2}_{-T/2} dt dt^\prime 
\frac{\left(X^i(t) - X^i(t^\prime)\right)^2}{(t-t^\prime)^2} \nn \\
&&  
+\frac{ieB_H}{2\hbar c} \int^{T/2}_{-T/2} dt \epsilon^{ij}
\p_t X^i X^j \nn\\
&& - \frac{V_0}{\hbar} \int^{T/2}_{-T/2} dt\left(
\cos \frac{2\pi X^1}{a}+ \cos \frac{2\pi X^2}{a} \right),
\eeq
where $i,j = 1, 2$. As in the case of the Schmid model, mapping 
the model to the string theory on a disk can be accomplished by
identifying the time as the boundary parameter $\sigma$ and 
scaling the field variables $X^i$:
\beq
S_{DHM} &=& \frac{1}{8\pi^2 \ap} \int^{\pi}_{-\pi} d\s d\s^\prime 
\frac{\left(X^i(\s) - X^i(\s^\prime)\right)^2}{(\s-\s^\prime)^2} \nn\\
&& + \frac{i\b}{4\pi} \int^{\pi}_{-\pi} 
d\s\epsilon^{ij} \p_\s X^i X^j \nn\\
&& - g \int^{\pi}_{-\pi} d\s \left(
\cos X^1+ \cos X^2 \right),
\eeq
where $2\pi\b = \frac{eB_H}{\hbar c}a^2$.
As in the case of the Schmid model, the action for the dissipative 
Hofstadter model can be interpreted as the boundary effective action 
for the string subject to a boundary periodic potential and the magnetic 
field on a disk with a boundary condition $X(\t=0,\s) = X(\s)$, 
\beq
&& \exp\left(-S_{DHM}\right) = \int D[X] \nn\\
&& \exp\Biggl[ - \frac{1}{4\pi \ap}
\int d\t d\s E_{ij} \left(\p_\t + \p_\s\right) X^i 
\left(\p_\t - \p_\s\right) X^j \nn\\
& & + \frac{g}{2} \int d\s \left(e^{iX^1}+ e^{-iX^1}+ e^{iX^2}+ 
e^{-iX^2} \right)\Biggr],
\eeq 
where $E_{ij} = \delta_{ij} + \frac{\b}{\a} \epsilon_{ij}$,  $\a = 
1/\ap$.
The relationship between the action for the dissipative 
Hofstadter model and the boundary state for the corresponding 
string theory is the same as before in Eq. (\ref{effective})
\beq
\langle 0 |B\rangle = \int D[X] \exp \left(-S_{DHM}[X] \right).
\eeq
Here, the boundary 
state $|B\rangle$ may be written as 
\beq \label{boundarystate}
|B\rangle &=& \exp\Biggl[g\pi \int_{\p M} \frac{d\s}{2\pi}\nn\\
&& ~~~ \left(
e^{iX^1}+ e^{-iX^1}+ e^{iX^2}+ e^{-iX^2} \right)\Biggr]|B_E\rangle 
\nn\\
&=& \sum_{n=0}^\infty \int d\s_1 \dots d\s_n 
\left(\frac{g}{2}\right)^n \frac{1}{n!} \sum_{{\bf q}_j=
\pm {\bf i}, \pm {\bf j}} \nn\\
&&~~~\prod_{j=1}^n \exp\left[i{\bf q}_j \cdot 
{\bf X}(\s_j)\right]|B_E\rangle.
\eeq
Integration over the zero mode imposes the constraints:
$\sum_i {\bf q}_i = 0$.
In the absence of the periodic potentials, the boundary state 
reduces to $|B_E\rangle$, which satisfies
\beq
\left(\delta_{ij} \p_\t X^j - \frac{\b}{\a}\epsilon_{ij} \p_\s X^j\right) 
|B_E\rangle =0. 
\eeq
Rewriting the boundary condition for $|B_E\rangle$ in 
terms of the oscillators 
\beq
\left(E_{ij} \a^j_{-n} + E^T_{ij} \ta^j_n\right) |B_E\rangle= 0, 
\quad p^i |B_E\rangle= 0,
\eeq
we find the explicit expression for $|B_E\rangle$ 
\beq
|B_E\rangle &=& \sqrt{\det E} \nn\\
&&\exp\left(-\sum_{n=1} \frac{1}{n}
\a^{i}_{-n}\left(E^{-1} E^T\right)_{ij} \tal^j_{-n} 
\right)|0\rangle, \nn\\
X^i(0, \sigma) &=& x^i + \o^i \s + \nn\\
&& ~~~~~ i \sqrt{\frac{\ap}{2}} \sum_{n\not= 0} 
\frac{1}{n} \left(\a^i_n - \ta^i_{-n}\right) e^{-in\s}. 
\eeq

The interaction term with the magnetic field may be completely removed 
by using the $O(2,2,R)$ transformation in the absence of the periodic 
potential; the boundary state has a simpler expression if we choose a new 
oscillator basis $\{\beta, \bar\beta\}$ which is related to the basis 
$\{\a, \bar\a\}$ by the $O(2,2,R)$ transformation 
\beq
\a^i_n = \left(G(E)^{-1}\right)^i{}_j \b^j_n,\quad
\tal^i_n = \left(G(E^T)^{-1}\right)^i{}_j \tbe^j_n.
\eeq
In terms of the new oscillator basis the boundary condition 
may be transcribed into the Neumann condition
\beq \label{neumann}
\left(\b^i_{-n} + \tilde \b^i_{n}\right) |B_E\rangle = 0.
\eeq
It is noteworthy that the oscillators $\{\beta, \bar\beta\}$ 
respect the worldsheet metric $G$ 
\beq
G = E^T E = \left(\begin{array}{cc}
  1 + \left(\frac{\b}{\a}\right)^2 & 0 \\
  0 &  1 + \left(\frac{\b}{\a}\right)^2 
\end{array} \right).
\eeq
Their commutation relations are
\beq
\left[\b^i_n, \b^j_m \right] &=& (G^{-1})^{ij} n 
\delta(n+m),\nn\\
\left[\tbe^i_n, \tbe^j_m \right] &=& (G^{-1})^{ij} n 
\delta(n+m) 
\eeq
and the boundary state $|B_E\rangle$ is rewritten as 
\beq
|B_E\rangle &=& \sqrt{\det E} \prod_{n=1} 
\exp\left(-\frac{1}{n} \b^i_{-n} G_{ij} \tbe^j_{-n}\right) \vert 0 
\rangle. 
\eeq
If we define a new string coordinate $Z$, 
\beq
Z^i(0, \sigma) &=& x^i + \o^i \s +\nn\\
&& \quad  i \sqrt{\frac{\ap}{2}} \sum_{n\not= 0} 
\frac{1}{n} \left(\b^i_n - \tb^i_{-n}\right) e^{-in\s},
\eeq
the relation between the two oscillator bases is summarized as 
\beq
X^i(\s) &=& X^i_L(\s)+X^i_R(\s) \nn \\
&=& \left(\delta^{ij}-\frac{\b}{\a}\e^{ij}\right) Z^j_L(\s) + \nn\\
&& ~~~~~~~ 
\left(\delta^{ij}+ \frac{\b}{\a}\e^{ij}\right) Z^j_R(\s).
\eeq

The new oscillator basis is also useful to analyze the boundary state 
$|B\rangle$ in Eq. (\ref{boundarystate}). Consider the product of vertex operators 
which appears in Eq. (\ref{boundarystate}) and rewrite it in terms of 
$\{\beta, \bar\beta\}$ 
\beq
&&\exp\left[i{\bf q}_1 \cdot {\bf X}(\s_1)\right] 
\exp\left[i{\bf q}_2 \cdot {\bf X}(\s_2)\right] = 
[{\rm Zero}~ {\rm Modes}] \nn\\
&&~ \exp\left[-\sqrt{\frac{\ap}{2}}\sum_{n\not=0} 
\frac{1}{n} q_{1k}(\b^k_n - \tb^k_{-n})e^{-in\s_1}\right] \nn\\
&&~ \exp\left[-\sqrt{\frac{\ap}{2}}\sum_{n\not=0} \frac{1}{n} 
q_{1k}(2\pi \ap B)^k{}_l(\b^l_n + \tb^l_{-n})e^{-in\s_1}\right]\nn\\
&&~ \exp\left[-\sqrt{\frac{\ap}{2}}\sum_{m\not=0}
\frac{1}{m} q_{2r}(\b^r_m - \tb^r_{-m})e^{-im\s_2}\right] \nn\\
&&~\exp\Biggl[-\sqrt{\frac{\ap}{2}}\sum_{m\not=0} \frac{1}{m} 
q_{2r}(2\pi \ap B)^r{}_s\nn\\
&& ~~~~~~~~~~~~~~~~~~~~~~~~~~~~~~~~(\b^s_n + 
\tb^s_{-n})e^{-in\s_2}\Biggr].
\eeq
Using the Baker-Hausdorff Lemma,
$e^A e^B = e^B e^A e^{[A,B]}$ where
\beq
A &=& -\sqrt{\frac{\ap}{2}}\sum_{n\not=0} \frac{1}{n} 
q_{1k}(2\pi \ap B)^k{}_l(\b^l_n + \tb^l_{-n})e^{-in\s_1},\nn \\ 
B &=& -\sqrt{\frac{\ap}{2}}\sum_{m\not=0}
\frac{1}{m} q_{2r}(\b^r_m - \tb^r_{-m})e^{-im\s_2},
\eeq
we have
\beq
[A,B] &=& -\ap {\bf q}_1 \cdot (2\pi\ap {\bf B} G^{-1}) \cdot {\bf q}_2 
\sum_{n\not=0} \frac{1}{n} e^{-in(\s_1-\s_2)} \nn\\
&=&  i \pi {\bf q}_1 \cdot (2\pi\ap^2 {\bf B}G^{-1}) \cdot {\bf q}_2\, 
{\rm sign}(\s_1-\s_2). 
\eeq
Note
\beq
-\sum_{n\not=0} \frac{1}{n} e^{-in(\s_1-
s_2)} e^{-\e|n|} &=& \ln \left[\frac{z_1-z_2 e^\e}{z_2 -z_1 e^\e} 
\cdot \frac{z_1}{z_2}\right] \nn\\
&=& i \pi \,{\rm sign}(\s_1-\s_2)
\eeq
where $z_j = e^{-i\s_j}$.
Thus, by repetition 
\beq
&&\prod_{j=1}^n \exp\left[i{\bf q}_j \cdot {\bf X}(\s_j)\right] 
|B_E\rangle \nn\\
&&~~~~~= \exp\left[i{\bf q}_n \cdot {\bf X}(\s_n)\right] 
\dots \exp\left[i{\bf q}_1 \cdot {\bf X}(\s_1)\right] 
|B_E\rangle \nn\\
&&~~~~~= \exp\left[\sum_{i>j}
i \pi {\bf q}_i \cdot (2\pi\ap^2 {\bf B}G^{-1}) \cdot {\bf q}_j\, 
{\rm sign}(\s_i-\s_j) \right] \nn\\
&& ~~~~~~~~~~
\prod_{j=1}^n \exp\left[i{\bf q}_j \cdot {\bf Z}(\s_j)\right]
|B_E\rangle,
\eeq
where the boundary condition in Eq. (\ref{neumann}) is used. 
Some algebra leads us to
\beq
&&\exp\left[\sum_{i>j}
i \pi {\bf q}_i \cdot (2\pi\ap^2 {\bf B}G^{-1}) \cdot {\bf q}_j\, 
{\rm sign}(\s_i-\s_j) \right] \nn\\
&&~~=
\exp\left[\sum_{i>j} i \pi 
\frac{\b}{\a^2+\b^2} \left(q^1_i q^2_j - q^2_i 
q^1_j\right)\, {\rm sign}(\s_i-\s_j)\right] \nn\\
&&~~= \exp\left[\sum_i 2\pi i \frac{\b}{\a^2+\b^2} 
q^1_i\left(\sum_{\s_i>\s_j} q^2_j\right)\right]. 
\eeq
Since $q^1_i, q^2_i = 0, \pm 1$ for $i=1,2$, if $\frac{\b}{\a^2+\b^2}$ is 
an integer, this phase due to the magnetic field reduces to 1. 
It simply implies that the periodic potential does not change under the 
mapping 
${\bf X} \rightarrow {\bf Z}$.
These circles 
\beq
\a^2+ \left(\b - \frac{1}{2n}\right)^2 = 
\left(\frac{1}{2n}\right)^2, \quad n \in Z
\eeq
on the two dimensional plane of $(\a,\b)$ may be called ``magic 
circles". 
Hence, on the magic circles, the dissipative Hofstadter model can be 
mapped into the string theory on a disk with the periodic 
potential only, {\it i.e.}, the boundary sine-Gordon model:
\beq
&&\exp\left(-S_{DHM}\right) = \int D[Z] \nn\\
&& \exp\Biggl[ - \frac{1}{4\pi \ap}
\int_M d\t d\s G_{ij} \left(\p_\t + \p_\s\right) Z^i 
\left(\p_\t - \p_\s\right) Z^j \nn\\
& & + \frac{g}{2} \int_{\p M} d\s \left(e^{iZ^1}+ e^{-iZ^1}+ e^{iZ^2}+ 
e^{-iZ^2} \right)\Biggr].
\eeq

It may be convenient to scale $\b^i, \tilde\b^i$ and $G$ such that
\beq
\b^i_n &&\rightarrow \sqrt{\ap} \b^i_n,\quad 
\tilde\b^i \rightarrow \sqrt{\ap} \tilde\b^i, \nn\\ 
G &&\rightarrow \frac{1}{\ap} G = \left(\begin{array}{cc}
 \frac{\a^2+\b^2}{\a} & 0 \\
  0 & \frac{\a^2+\b^2}{\a} 
\end{array} \right).
\eeq
The points where the effective worldsheet metric becomes a unit 
metric form a circle called the ``critical circle":
\beq
\frac{\a^2+\b^2}{\a} = 1.
\eeq
On the critical circle, the commutation relations 
between $\b^i_n$ and $\tilde\b^i_n$ (after scaling) 
respect the unit metric
\beq
\left[\b^i_n, \b^j_m \right] &=& \delta^{ij} n 
\delta(n+m),\nn\\
\left[\tbe^i_n, \tbe^j_m \right] &=& \delta^{ij} n 
\delta(n+m).
\eeq
Accordingly, the boundary state $|B_E\rangle$ may be written on the 
critical circle as
\beq
|B_E\rangle &=& \sqrt{\det E} \prod_{n=1} 
\exp\left(-\frac{1}{n} \b^i_{-n}  \tbe^i_{-n}\right) \vert 0 
\rangle. 
\eeq
The points where the magic circles meet the critical circle are
termed magic points; at the magic points, the dissipative 
Hofstadter model is equivalent to a set of two independent 
critical boundary sine-Gordon models:
\beq
&& \exp\left(-S_{DHM}\right) \nn\\
&&= \int D[Z] \exp\Biggl[ - \frac{1}{4\pi}
\int_M d\t d\s \left(\p_\t + \p_\s\right) Z^i 
\left(\p_\t - \p_\s\right) Z^i \nn\\
& & ~~+ \frac{g}{2} \int_{\p M} d\s \left(e^{iZ^1}+ e^{-iZ^1}+ e^{iZ^2}+ 
e^{-iZ^2} \right)\Biggr].
\eeq
In other words, the system can be described by a set of two decoupled 
Schmid models at the critical points, which can be mapped into the free fermion theory.

\begin{center}
{\bf IV. CONCLUSIONS}
\end{center}

We discuss the dissipative Hofstadter model in the framework of 
the boundary state formulation, which has been developed for the 
string theory. As pointed out by Callan and Thorlacius 
\cite{callan90}, if we map the quantum system of the Hofstadter 
model onto the boundary of the string worldsheet disk, we see
that the Caldeira-Leggett coupling, responsible for the 
dissipation, is already built in; the bulk string degrees of freedom on the disk play the role of the bath. Therefore, the boundary state 
formulation, which is the most convenient tool to discuss the string 
theory on a disk, could be the most effective framework for the 
dissipative Hofstadter model. This point has been elaborated by
Callan and Thorlacius \cite{callan90}, who dealt with the Schmid model 
only, and their work has been further extended to the dissipative 
Hofstadter model by Callan {\it et al.} \cite{callan95}. However,
the full advantage of the boundary state formulation has not
been taken in the previous work. 

One of the most useful tools for the 
two-dimensional theories is fermionization (or bosonization).
This is particularly useful when the bosonic (fermionic) theory becomes exactly 
solvable if it is fermionized (bosonized). The Schmid model and 
the dissipative Hofstadter model come under this category.
Combining the boundary state formulation and the fermionization 
technique, in this paper we show that the Schmid model at the critical
point, being equivalent to the critical sine-Gordon model, can be exactly
solvable. The exact boundary state for the Schmid model at the critical 
point is nothing but the fermion boundary state for the rolling tachyon
constructed in ref. \cite{Hassel}. Thus, using the exact boundary state,
one can calculate any physical quantity exactly.
To illustrate, let us calculate the mobility,
which determines if the ground state of the system is localized or 
delocalized. The mobility is a two point correlation function,
\beq
M(\s,\sp) &=& \langle \p_\s X(\s) \p_\s X(\sp) \rangle \nn\\ 
&=& \langle 0| \p_\s X(\s) \p_\s X(\sp) |B \rangle. 
\eeq
The exact boundary state and the fermionization technique enable us to 
calculate the mobility exactly:
\beq
M(\s,\sp) &=& \langle 0|\left(J^3_L(\s)+ J^3_R(\s)\right)
\left(J^3_L(\sp)+ J^3_R(\sp)\right) \nn\\
&& \prod_{r=\half} 
\exp\left[\psi^{\dagger}_{-r} U^{-1} i\s^1 \tpsi_{-r} - \tpsi^{\dagger}_{-r} 
i\s^1 U \psi_{-r}\right]|0\rangle \nn\\
&=& - \frac{1}{2} (1-\pi^2 g^2) 
\sin^{-2}\frac{(\s-\sp)}{2},
\eeq
where $|g| \le 1/\pi$. (Note that the Ramond sector 
contains only the states with 
half integer momenta. This implies that the vacuum state belongs
to the NS-sector. Thus, there is no contribution from the 
Ramond sector.)

Using the boundary state formulation and the $O(2,2,{\bf R})$ 
transformation of the string theory, we also show that the 
dissipative Hofstadter model on the magic circles, where  
flux/unit cell $\b$ is an integer multiple of
the friction/unit cell $\a$, can be mapped into the Schmid model.
Hence, the equivalence between the two models is made manifest.
The equivalence between the dissipative Hofstadter model on the magic 
circles and the Schmid model has been discussed only in the calculation of the partition function \cite{freed93}. 
The exact calculation of the mobility of the Schmid model at the 
critical point can be easily extended to the case of the 
dissipative Hofstadter model; at the magic points the mobility of the
dissipative Hofstadter model can be calculated by using the exact 
fermion boundary state as follows:
\beq
&& M^{ij}(\s,\sp) \nn\\
&&= \langle \p_\s X^i(\s) \p_\s X^j(\sp) \rangle \nn\\
&&= \langle 0| \p_\s X^i(\s) \p_\s X^j(\sp) |B \rangle \nn\\
&&= \langle 0| \Biggl[\left(\delta^{ik} - \frac{\b}{\a} \e^{ik}\right) 
\p_\s Z^k_L(\s) \nn\\
&&~~~~~~~~~~~~~~~~ + \left(\delta^{ik} + \frac{\b}{\a} \e^{ik}\right) 
\p_\s Z^k_R(\s) \Biggr] \nn\\
&& ~~~~~ \Biggl[\left(\delta^{jl} - \frac{\b}{\a} \e^{jl}\right) 
\p_\s Z^l_L(\sp) \nn\\
&&~~~~~~~~~~~~~~~~ + \left(\delta^{jl} + \frac{\b}{\a} \e^{jl}\right) 
\p_\s Z^l_R(\sp) \Biggr]
|B \rangle,
\eeq
where 
\beq
|B\rangle = |BD\rangle_1 \otimes |BD\rangle_2.
\eeq
Since $Z^k_{L/R}$ can be represented as the fermion current 
operators $J^{3k}_{L/R}$, the exact calculation of the mobility 
is not a difficult task. The result of the calculation and 
a more detailed discussion on the dissipative Hofstadter model
will be saved for future work.

\begin{acknowledgments}
The work of TL was done during his visit to ICTP (Italy), KIAS (Korea) 
and PITP (Canada) and was supported by KOSEF through the Center 
for Quantum Space-Time (CQUeST). TL thanks G. Semenoff and P. Stamp 
for useful discussions during the early stage of this work. 
The work of JK was supported by the
Ministry of Science and Technology of Korea through the Creative 
Research Initiative.
\end{acknowledgments}



\end{document}